\documentclass[amsmath,twocolumn,showpacs,aps,prl]{revtex4}
\usepackage{bm}
\usepackage{graphics}
\usepackage{graphicx}

\begin{document}

\title{Anomalous transition temperature oscillations in LOFF state}

\author{A. A. Zyuzin and A. Yu. Zyuzin}

\affiliation{A.F. Ioffe Physico-Technical Institute of Russian
Academy of Sciences, 194021 St. Petersburg, Russia}


\begin{abstract}
We consider Aharonov - Bohm effect at normal metal-inhomogeneous
LOFF superconducting state transition. It is shown that magnetic
flux can increase the transition temperature and AB oscillations
can have the double-peak structure at one period. Expressions for
fluctuational heat capacity and persistent current are calculated
for a thin ring and a cylinder. We also discuss the effect of
fluctuations interaction in the nonuniform states in the vicinity
of the superconducting transition.
\end{abstract}
\maketitle

\section{Introduction}
Spin polarization of the Cooper pairs in magnetic field destroys
the superconducting state at Chandrasekhar- Clogstone limit when
paramagnetic energy coincides with the superconducting
condensation energy. Paramagnetic limit is attained at critical
field $H_{p}=\sqrt{2}\Delta/2\mu_B$ \cite{bib: CC}, where $\Delta$
is the superconducting gap, $\mu_B$ is the Bohr magneton. Orbital pair breaking effect usually
dominates over the paramagnetic limit. However, orbital effect
could be suppressed in low dimensional systems (thin wires) or by
applying magnetic field parallel to the conductive planes of
quasi- two dimensional systems.

Spin polarization in external magnetic field or intrinsic
exchange fields could lead to the formation of inhomogeneous
superconductivity. Larkin and Ovchinnikov \cite{bib: LO}, Fulde
and Ferrel \cite{bib: FF} predicted the existence of the
nonuniform superconducting state in a ferromagnetic
superconductors at low temperatures in magnetic field larger than
critical $H_p$ (see for a review \cite{bib: LOFF1, bib: LOFF2}).

LOFF state is formed by Cooper pairs with nonzero momentum $\sim
2\mu_{B}H/v_F$, where $v_F$ is the Fermi velocity, and at fields
higher than the paramagnetic limit has lower
energy compared to the uniform superconducting state . This finite momentum of Cooper
pairs results in a spatial modulation of the superconducting order
parameter.

Mathematically, LOFF state appears due to the change in sign of
coefficient $\beta$ at the gradient term of the Ginzburg - Landau
(GL) free energy functional $\beta |\mathbf{\nabla}\Psi|^2$, where
$\Psi$ is the order parameter. Coefficient $\beta$ is a function
of temperature and Zeeman energy $\mu_B H$. In BCS model it
becomes negative at high magnetic fields $H>1.07 T_c(0)/\mu_B$ and
low temperatures $T < 0.56 T_c(0)$, where $T_c(0)$ is the
temperature of transition to superconducting state at zero magnetic field, signaling of the formation of
nonuniform LOFF state. As a result one has to take into account
higher terms in the GL functional expansion $
|\mathbf{\nabla}^2\Psi|^2$.

The theoretical research of nonuniform LOFF state includes, for
example, the study of impurities effect \cite{bib: Aslamazov} that
suppresses the region of nonuniform superconductivity, the
LOFF-like proximity effect at the ferromagnetic - superconductor
boundary \cite{bib: BuzdinFerro}, interplay of orbital and
paramagnetic effects \cite{bib: HouzetMineev, bib: HouzetBuzdin,
bib: Arg}, the study of phase transitions in the vicinity of the
tricritical point \cite{bib: Kachkachi}, intrinsic pinning of
vortices in layered superconductors \cite{bib: Bulaevskii}.

Heavy- fermion compound $\mathrm{CeCoIn_5}$ was found to show the
signatures of LOFF phase. The existence of the LOFF state in
heavy- fermion superconductor was experimentally investigated by
specific heat measurements \cite{bib: HF1, bib: HF3, bib: HF4} and
nuclear magnetic resonance \cite{bib: HF2, bib: HF5}.

The phase transition between possibly the LOFF state and the
homogenous superconducting state was reported for organic
superconductors such as $\mathrm{\lambda-(BETS)_2 FeCl_4}$
\cite{bib: OS1, bib: OS2, bib: OS3, bib: OS4} with quasi-2D
electronic structures and organic $\mathrm{(TMTSF)_2ClO_4}$
\cite{bib: OS5} with quasi-1D electronic structure.

These experiments were focused on the identification of the phase
transition inferred from a kink of thermal conductivity \cite{bib:
OS3}, observation of peculiar properties -dip structures- in the
resistance \cite{bib: OS2} and changes in the rigidity of the
vortex system \cite{bib: OS4}. The thermodynamic evidence of the
existence of narrow intermediate state attributed to LOFF state, which
separates the uniform superconducting state and normal state based
on specific heat measurements was presented in paper \cite{bib:
OS1}. Finally, the transition temperature dependence on the strength
and direction of the magnetic field was studied in paper
\cite{bib: OS5}

Nonuniform state of condensate was uncovered in systems of
ultra-cold atoms in optical lattices \cite{bib: OLattice}.
Experimentally optical lattices could be formed by standing wave
laser which provides a periodical potential for ultra cold atomic
gas. Recently, the observation of phase transition between normal
and nonuniform superfluid state was reported for systems of
strongly interacting Fermi-gas with imbalanced spin population
\cite{bib: OLattice2}.

Crossovers between different fluctuational regimes of
paraconductivity and specific heat in the vicinity of the LOFF
transition were discussed theoretically in paper \cite{bib:
LOFF_Fluct}. Authors showed that these fluctuational contributions
have specific temperature dependencies compared to the case of
uniform superconductivity and could serve as an additional
indicator of the LOFF state.

In present paper we consider hollow superconducting cylinder/ring
threaded by magnetic flux (fig. \ref{fig: 1}). We calculate the
expressions for the persistent current in thin superconducting
ring and present numerical results for specific heat and
persistent current for the cylinder. Magnetic flux dependence of
the current demonstrates the double-peak structure in Aharonov
-Bohm (AB) oscillations. We also study the effect of fluctuations
interaction on the nonuniform states in low- dimensional
inhomogeneous superconductors. This research is motivated by the
fact that the fluctuation region in nonuniform systems is much
larger than in the case of uniform states and requires separate
theoretical study.

\section{GL free energy}

\begin{figure}[t] \centering
\includegraphics[width=7.5cm]{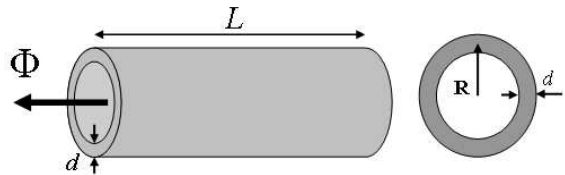}\\
\caption{Thin superconducting cylinder and rind}\label{fig: 1}
\end{figure}

We consider AB oscillations in quasi one dimensional ring and thin
-walled cylinder which transverse size $d$ is much smaller than
the radius $R$, see fig. (\ref{fig: 1}). In case of second order
normal metal-LOFF transition the Ginzburg- Landau free energy
functional above transition temperature could be written as
\begin{equation} \label{FreeEnergy}
F=\int d \mathbf{r}\left(a(T-\widetilde{T}_c)|\Psi|^2 + \beta
|\mathbf{D}\Psi|^2 +\delta |\mathbf{D}^2 \Psi|^2\right)
\end{equation}
where $\beta = -|\beta|$ and $\mathbf{D}=-i\mathbf{\nabla} -
(2e/c)\mathbf{A}$, while tangent component of the vector potential
is given as $A_{\varphi}=\Phi/2\pi R$. Representing the order
parameter as
\begin{equation}\label{psi}
\Psi(\mathbf{r}) = \sum_{n, k} \Psi_{n}(k) e^{i\varphi n} e^{ikz}
\end{equation}
where z is coordinate along the cylinder. We can write the free
energy functional as
\begin{equation}
F= V \sum_{n,k} E_n(k) |\Psi_{n}(k)|^2
\end{equation}
where $V$ is volume of the sample and
\begin{eqnarray}\label{cyl-energy} \nonumber
E_n(k)& =& a(T-T_{c} ) +\\
&+&\frac{|\beta|}{2Q^2 R^4}\left( (n-\frac{\Phi}{\Phi_0})^2
+(kR)^2-(QR)^2\right)^2
\end{eqnarray}
Here $\Phi_0 = e / \pi c$ is the flux quantum,
\begin{equation}
Q=\sqrt{\frac{|\beta|}{2\delta}}
\end{equation}
is the modulus of superconducting modulation wave-vector in
inhomogeneous LOFF state, and
\begin{equation}\label{high-T}
T_{c}=\widetilde{T}_c+\frac{\beta^2}{4a\delta}
\end{equation}
is the transition temperature at $R\rightarrow\infty$.
Recently \cite{bib: ours}, we have examined the Aharonov - Bohm
oscillations in thin ring near LOFF - metal transition. In
contrast to the uniform superconductivity the applied magnetic
flux can increase the transition temperature of LOFF state and AB
oscillations could have double peak structure. The nature of these
effects is the fluctuation energy spectrum of the inhomogeneous
state. To see it in more detail let us consider the ring threaded
by magnetic flux $\Phi$. In this case spectrum $E_n(k)$ is given by eq.
(\ref{cyl-energy}) at $k=0$
\begin{equation}\label{first_energy}
E_n = a(T-T_{c}) + \frac{|\beta|}{2Q^2 R^4}\left(
(n-\frac{\Phi}{\Phi_0})^2 -\phi^2 \right)^2
\end{equation}
where we introduce $\phi=QR$.

The independence of spectrum on $k$ allows one to introduce flux
dependent transition temperature, corresponding to $E_n=0$ as
\begin{equation}\label{crit-temp}
T_{c}(\Phi)=T_{c}-\frac{|\beta|}{2aQ^2 R^4}\min\left(
(n-\frac{\Phi}{\Phi_0})^2 -\phi^2 \right)^2
\end{equation}

The transition temperature into the LOFF state of the
superconducting ring is defined by $n_{\pm}$ which are nearest
integers to the corresponding values $\Phi/\Phi_0 \pm \phi$.

Generally, integers $n_{\pm}$ do not correspond to the minimum
of the energy (\ref{first_energy}) which provides the highest
transition temperature (\ref{high-T}) of the system. Thus, one can
both increase or decrease $E_n$ and correspondingly the transition
temperature $T_{c}(\Phi)$ by changing the radius of the ring or by
applying magnetic flux $\Phi$. This is in contrast to the case of
metal-uniform superconductor transition where transition
temperature $T_{c}(\Phi)$ always decreases with applying magnetic
flux. Moreover, the degeneracy of the energy spectrum allows
the system to jump between $E_{n_{+}}$ and $E_{n_{-}}$ states
leading to peculiar properties of AB effect such as double-peak
structure per period of oscillations.

\begin{figure}[t] \centering
\includegraphics[width=7cm]{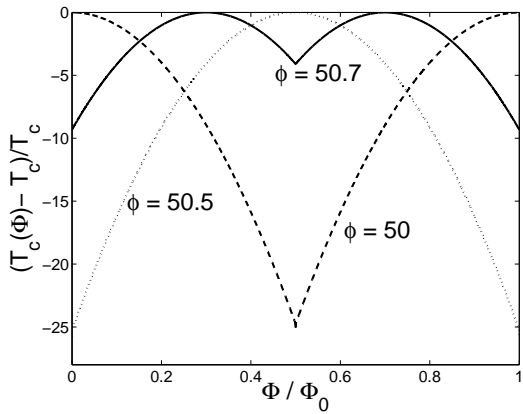}
\caption{One period of $(T_c(\Phi)-T_c)/T_c$ oscillations in
magnetic flux for a set of $\phi = QR = (50,50.5,50.7)$, where
$|\beta|/2aQ^2 R^4= 0.01$. See eq. (\ref{crit-temp}) in the text.
}\label{fig:2}
\end{figure}
Let us discuss the possible temperature dependencies shown in fig.
(\ref{fig:2}) for the case of large $\phi \gg 1$. In this regime
the transition temperature behavior is given by
\begin{equation}
(T_c(\Phi)-T_c)/T_c = - \frac{2|\beta|}{aT_{c}R^2}\max
(f_{+}(\Phi)^2 , f_{-}(\Phi)^2 )
\end{equation}
Where $f_{\pm}(\Phi)$ is the distance between $\Phi/\Phi_0 \pm
\phi$ and corresponding integer $n_{\pm}$.

If $0 < {\phi} < 0.5$ then for applied magnetic flux in the range
$0<\Phi < \Phi_0/2$ the transition temperature behavior is
governed by $f_{-}$. When $\Phi$ reaches $\Phi_0/2$ the crossover
from $f_{-}$ to $f_{+}$ takes place and the further magnetic flux
dependence is described by $f_{+}$ leading to the double-peak
structure of oscillations. Note that if $0.5 < {\phi} < 1$ one has
the opposite crossover from $f_{+}$ to $f_{-}$.

Suppose the value of phase $\phi$ equals to the half an integer
number (dotted line in fig. (\ref{fig:2})). Transition temperature
first decreases as $(T_c(\Phi)-T_c)/T_c \propto -(\Phi/\Phi_0)^2$
with increasing applied magnetic flux. The maximal depression of
$T_c(\Phi)$ occurs when $\Phi/\Phi_0 = 1/2$ and has a value of
$(T_c(\Phi)-T_c)/T_c=-\frac{|\beta|}{2aT_{c}R^2}$. Further
increase of $\Phi$ leads to the increase of transition temperature
as $(T_c(\Phi)-T_c)/T_c \propto (\Phi/\Phi_0)^2 - 1 $. Here both
$f_{+}$ and $f_{-}$ give equal dependence on $\Phi$.

Finally, if the value of $\phi$ equals to the integer number then
one has the opposite case (dashed line in fig.\ref{fig:2}).
Transition temperature increases with applied magnetic flux
starting from the value $(T_c(\Phi)-T_c)/T_c =
-\frac{|\beta|}{2aT_{c}R^2}$. When $\Phi$ reaches $\Phi_0/2$, the
crossover between $f_{+}$ and $f_{-}$ leads to further decrease of
$(T_c(\Phi)-T_c)/T_c$.

The variations of $T_c(\Phi)$ quantitatively explain flux
dependence of physical quantities of quasi one dimensional ring.
Interestingly, summation over momentum $k$ in case of cylinder
does not wash out the peculiarities of flux dependence. In this case
$kR/2\pi$ plays a role of random phase and the cylinder could be
considered as a set of rings with different phases $\phi$.
Superposition of different types of oscillations as we will show
below leads to fact that the oscillation peculiarities appear at
lower temperatures and/or smaller radiuses of the ring.

\section{Specific heat}
Carrying out the integral over the real and imaginary parts of
order parameter $\Psi_n(k)$ one obtains the expression for
fluctuational part of thermodynamical potential
\begin{equation}\label{energy}
\Omega = T \sum_{n,k} \ln\left(\frac{E_n(k)}{\pi T}\right)
\end{equation}
We accept units where $k_B =1$. Fluctuational correction to
specific heat is given as
\begin{equation}\nonumber
C= -\frac{T}{V}\frac{\partial^2 \Omega}{\partial T^2}
\end{equation}
where $V$ is the volume of the sample. Taking derivatives over
the temperature dependence of $E_n(k)$ one obtains \cite{bib:
Larkin_Varlamov}
\begin{equation}
C= \frac{(aT_c)^2}{V} \sum_{n,k}E_n^{-2}(k)
\end{equation}
Performing the Poisson transformation one obtains the fluctuation
specific heat of the superconducting cylinder at temperatures
$T>T_c$
\begin{equation}
C= \frac{q}{VR(T/T_c - 1)^{2}} \Re \sum_{m,k}\int\frac{e^{2\pi
im(\Phi/\Phi_0 +tq)}dt}{[1+(t^2-z + (kR/q)^2)^2]^2}
\end{equation}
where we introduce parameters
\begin{eqnarray}\nonumber
q &=& R\sqrt{\frac{\sqrt{2}Q}{\zeta}} \\
z &=& \frac{Q\zeta}{\sqrt{2}}
\end{eqnarray}
Parameter $z$ characterizes the LOFF inhomogeneity of the
fluctuations and is proportional to the number of modulations of
superconducting fluctuations of correlation length $\zeta$.

Correlation length $\zeta$ measures the  scale of superconducting
fluctuations and is defined as
\begin{equation}\label{coh1}
\zeta =\sqrt{\frac{|\beta|}{a(T-T_{c} )}}
\end{equation}

Now one can integrate over $t$ and make summation over $m$. The expression for the specific heat is then
\begin{equation}\label{sp-heat}
C = \frac{\pi q}{4VR(T/T_c - 1)^{2}} \Re \sum_{k} \left[ f(\Phi) +
f(-\Phi) \right]
\end{equation}
Here
\begin{equation}
f(\Phi) = \left[\frac{1}{g}+ \frac{i/2}{g^3}\right]
\frac{1+e^{2\pi i\varphi }} {1- e^{2\pi i\varphi } }
+\left[\frac{2\pi q}{g^2}\right]\frac{e^{2\pi i\varphi}}{1-e^{2\pi
i\varphi}}
\end{equation}
while $\varphi = \Phi/\Phi_0 +qg$ and $g = (z-(kR/q)^2 +i)^{1/2}$.
\begin{figure}[t] \centering
\includegraphics[width=8cm]{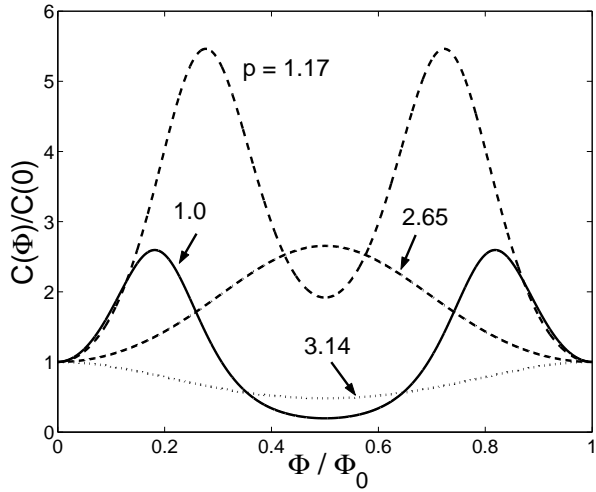} \caption{Magnetic flux dependence
of specific heat of the ring for $Q\zeta/\sqrt{2} = 10$. Parameter
$ p = \sqrt{2}\pi R/\zeta = [1.0, 1.17, 2.65, 3.14]$ measures the
ratio of rings radius to the correlation length } \label{fig:3}
\end{figure}

\begin{figure}[t] \centering
\includegraphics[width=8cm]{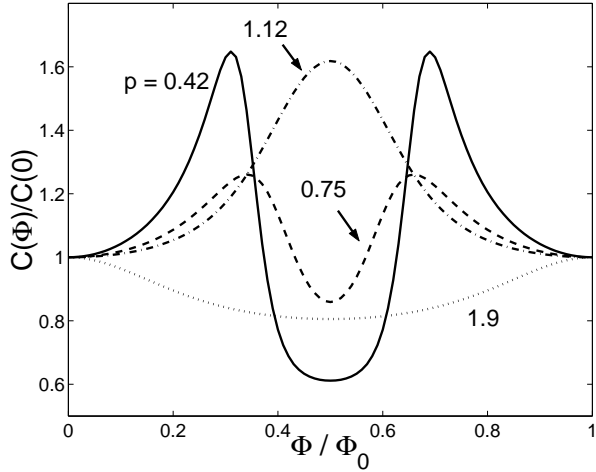} \caption{Magnetic flux dependence
of specific heat of the cylinder for $Q\zeta/\sqrt{2} = 10$.
Parameter $ p = \sqrt{2}\pi R/\zeta = [0.42, 0.75, 1.12, 1.9]$
measures the ratio of rings radius to the correlation length }
\label{fig:4}
\end{figure}

Specific heat of the ring is determined by $k=0$ term in
expression (\ref{sp-heat}). The detailed analysis of the specific
heat magnetic flux dependence for the case of thin superconducting
ring was given in the paper \cite{bib: ours}. There it was shown
that magnetic flux can increase the critical temperature of LOFF
state and AB oscillations could have double peak structure.

Here in fig.\ref{fig:3} we present the typical magnetic flux
dependencies of the specific heat of the ring. It is seen that for
$Q\zeta >>1$, where number of modulations of superconducting
fluctuations is large, specific heat exhibits pronounced double
peak structure when  $ \sqrt{2}\pi R \sim \zeta$.  In the case of
small scale superconducting fluctuations when $\sqrt{2}\pi R >>
\zeta$ one obtains the random sign behavior of specific heat on
magnetic flux. In this regime magnetic flux can both increase or
decrease the specific heat.

Summation over momentum in the case of cylinder is defined as
\begin{equation}\nonumber
\sum_{k}=L\int \frac{dk}{2\pi}
\end{equation}
where L is length of cylinder.

Numerical results for thin superconducting cylinder are shown on
fig.\ref{fig:4}. It is seen from the fig.\ref{fig:4} that magnetic
flux can also either increase or decrease specific heat and leads
to the double-peak structure in oscillations. However, the
transition to the double-peak structure regime appears at smaller
radiuses of the cylinder compared to the radius of ring. This is
the consequence of the averaging procedure over momentum $k$.

\section{Persistent current}
In this section we will discuss the persistent current in AB
effect. Expression for the persistent current is given as
\begin{equation}
I = - \frac{\partial \Omega}{\partial \Phi} =
-T\sum_{n,k}\frac{\partial E_n(k)/\partial \Phi}{E_n(k)}
\end{equation}
Using equation (\ref{energy}) and performing the Poisson
transformation we obtain at $T>T_{c}$
\begin{equation}\label{PC}
I = \frac{2T}{\Phi_0}\sum_{m, k}\int dt
\frac{2t(t^2+y^2-z)}{1+(t^2+y^2-z)^2}e^{2\pi im( \Phi/\Phi_0+ tq)}
\end{equation}
Performing summation over $m$ and integration over $t$ we conclude with
\begin{equation}\label{per-curr}
I = -\frac{4\pi T}{\Phi_0}\Re\sum_k
\frac{\sin{(2\pi\Phi/\Phi_0)}}{\cos{(2\pi\sqrt{\phi^2-(Rk)^2+iq^2})}-\cos{(2\pi\Phi/\Phi_0)}}
\end{equation}

Persistent current of the ring is determined by term $k=0$ in
expression (\ref{per-curr}).

\subsection{The regime of strong inhomogeneity $Q\zeta >>1$}
\begin{figure}[t] \centering
\includegraphics[width=8cm]{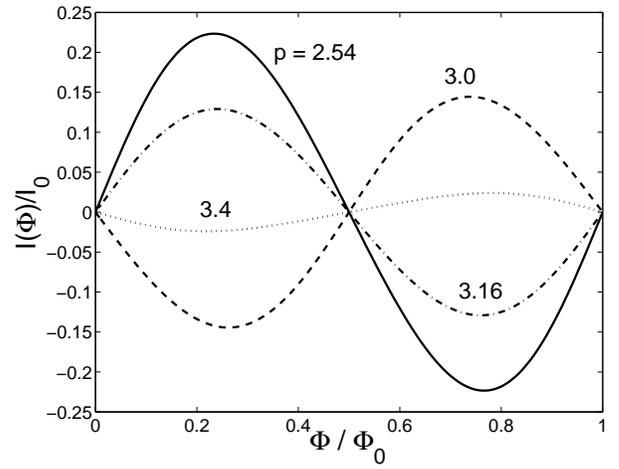} \caption{Magnetic flux dependence
of persistent current of the cylinder measured in units $I_0 = 2
T_c L /R \Phi_0$ for $Q\zeta/\sqrt{2} = 10$ and large radius. Here
$ p= \sqrt{2}\pi R/\zeta = [2.54, 3.0, 3.16, 3.4]$ } \label{fig:6}
\end{figure}

Let us first consider the temperature regime in the vicinity of
the LOFF-metal transition which corresponds to the large number of
modulations of superconducting fluctuations $Q\zeta
>>1$. We first suggest the radius of the superconducting
ring/cylinder being larger than the correlation length
\begin{equation}\nonumber
R\gg \zeta
\end{equation}
\begin{figure}[t] \centering
\includegraphics[width=8cm]{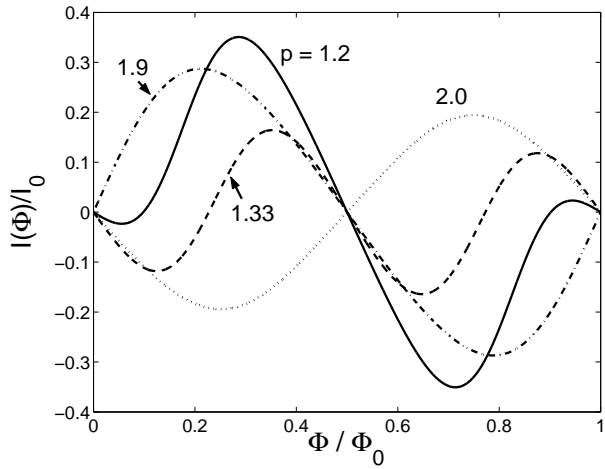} \caption{Persistent current of the
ring measured in units $I_0 = 4\pi T_c /\Phi_0$ for
$Q\zeta/\sqrt{2} = 10$ as a function of magnetic flux. Parameter
$p= \sqrt{2}\pi R/\zeta = [1.2, 1.2, 1.33, 2.0]$ } \label{fig:5}
\end{figure}
To calculate the persistent current for the superconducting ring
in this regime one has to take into account mode $m=1$ in
eq.(\ref{PC}) since higher modes will be exponentially suppressed. As
a result, the persistent current in the ring can be estimated as
\begin{equation}\label{pc_largeRing2}
I\simeq - \frac{8\pi T}{\Phi_0}\cos{(2\pi
\phi)}\sin{(2\pi\Phi/\Phi_0)e^{-\sqrt{2}\pi R /\zeta}}
\end{equation}
Depending on the sign of the random phase factor $\cos(2\pi\phi)$
persistent current could produce either diamagnetic or
paramagnetic response at small flux: fig.\ref{fig:5}. That is in
contrast to the case of homogenous superconductor- metal
transition. The numerical result for the cylinder geometry is
presented in the fig.\ref{fig:6}. One sees that the magnetic flux
dependence of the persistent current is also sensitive to the
random phase.

\begin{figure}[t] \centering
\includegraphics[width=8cm]{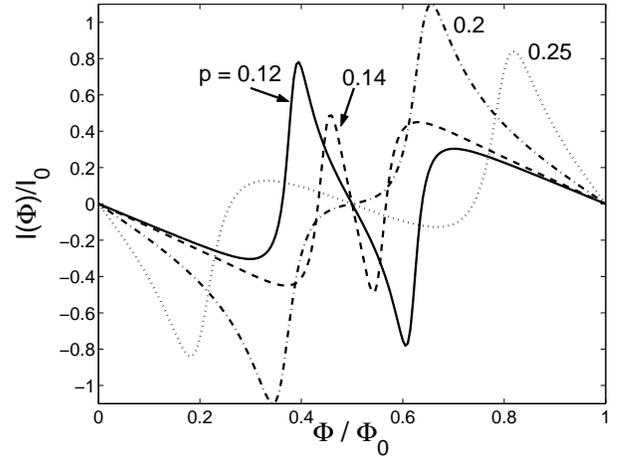} \caption{Magnetic flux dependence
of persistent current of the cylinder measured in units $I_0 = 2
T_c L /R \Phi_0$ for $Q\zeta/\sqrt{2} = 10$ and small radius. Here
$ p= \sqrt{2}\pi R/\zeta = [0.12, 0.14, 0.2, 0.25]$ }
\label{fig:7}
\end{figure}

Now let us discuss the regime of nonuniform superconductivity in
ring/cylinder with small radius
\begin{equation}\nonumber
\zeta >R
\end{equation}
Again we concentrate on the temperatures in the vicinity of the
LOFF-metal transition. One obtains for the current in thin
superconducting ring
\begin{equation}\label{pc_smallRing}
I= - \frac{2\pi T}{\Phi_0}[f(\Phi)-f(-\Phi)]
\end{equation}
where
\begin{equation}
f(\Phi) \simeq \frac{ 2\sin{(2\pi(\phi+\Phi/\Phi_0))}} {1+(\pi
R/\zeta)^2 - \cos{(2\pi(\phi+ \Phi/\Phi_0))}}
\end{equation}
In this case one observes the pronounced double - peak structure
of the persistent current oscillations in thin superconducting
ring: fig. (\ref{fig:5}).
The same result also holds for the superconducting cylinder and
the numerical calculations of the current dependence on the
magnetic flux are presented in the fig.\ref{fig:7}.
Again, the double-peak oscillations structure exhibits at smaller
radiuses of the cylinder compared to the ring due to summation
over momentum $k$.

It is of interest to compare the result obtained above with the
case of homogenous superconductor -normal metal transition. In
this regime the persistent current in the thin ring is given by the
expression \cite{bib: PC1, bib: PC2}
\begin{equation}\label{per-cur-un1}
I = -\frac{2\pi T}{\Phi_0}
\frac{\sin{(2\pi\Phi/\Phi_0)}}{\cosh{(2\pi R/\zeta)} -
\cos{(2\pi\Phi/\Phi_0)}}
\end{equation}
Thus if the radius of the ring is larger than the coherence length
then
\begin{equation}\label{per-cur-un2}
I \simeq -\frac{4\pi T}{\Phi_0} e^{-2\pi
R/\zeta}\sin{(2\pi\Phi/\Phi_0)}
\end{equation}

Comparing expressions (\ref{pc_largeRing2}) and
(\ref{pc_smallRing}) with expressions (\ref{per-cur-un1}) and
(\ref{per-cur-un2}) we see that the persistent current in
nonuniform case is the result of the summation of two usual
currents with phases shifted by $\pm \phi$.

\subsection{The regime of weak inhomogeneity $Q\zeta <1$}

Finally, we consider the oscillation regime in the vicinity of
metal - LOFF transition where $Q\zeta <1$. This regime corresponds
to weakly inhomogeneous superconducting fluctuations when $\beta
\rightarrow 0$. The number of LOFF modulations per correlation
length $\zeta$ is small. In equation (\ref{per-curr}) for
persistent current we will also consider the following condition
\begin{equation}
QR > \zeta/R
\end{equation}
This conditions implies that if the radius of the ring is larger
than the correlation length the number of LOFF modulations per
circumference of the ring $QR$ should be large. This condition
can be rewritten as
\begin{equation}
R > \xi
\end{equation}
where now effective coherence length is given as
\begin{equation}\label{coh2}
\xi =\left(\frac{\zeta}{\sqrt{2} Q}\right)^{1/2}
\end{equation}

\begin{figure}[t] \centering
\includegraphics[width=8.5cm]{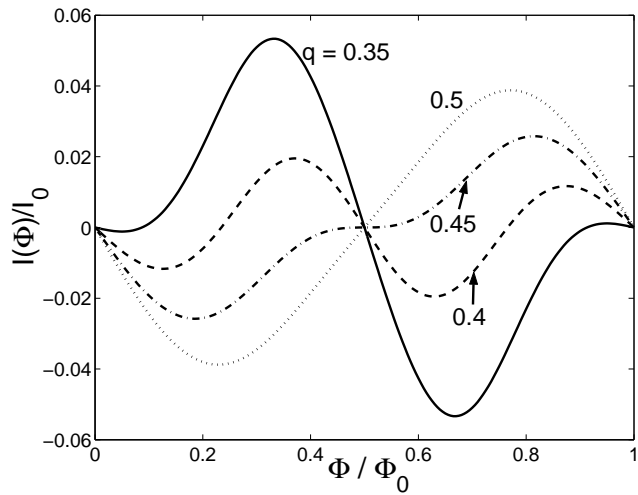} \caption{Persistent current of the
cylinder as a function of magnetic flux measured in units $I_0 = 2
T_c L /R \Phi_0$. For the case of $Q\zeta/\sqrt{2} = 0.1$ and $q=
R\sqrt{\sqrt{2}Q/\zeta} = [0.35, 0.4, 0.45, 0.5]$} \label{fig:8}
\end{figure}

With these assumptions one concludes with the following expression
for the persistent current of the thin ring
\begin{equation} \label{FarRing}
I= - \frac{16\pi T}{\Phi_0}\sin{\frac{2\pi\Phi}{\Phi_0}}
\cos{(\sqrt{2}\pi R/ \xi)}e^{-\sqrt{2}\pi R/\xi}
\end{equation}
This result is illustrated in the fig. (\ref{fig:8}), where the
persistent current dependence on the magnetic flux is presented.
One sees that the amplitude of the oscillations decreases compared
to the temperatures regime in strongly inhomogeneous fluctuations
$Q\zeta > 1$.

\section{applicability of gaussian approximation}
Here we will examine applicability of gaussian approximation in
the vicinity of LOFF - metal transition in general. A first step
of estimating the fluctuation interaction correction above $T_c$
is to take into account the contribution of neglected $|\Psi |^4$
and $|\Psi|^6$ terms. Last term is important in case of first
order LOFF-normal metal transition. GL functional is then
\cite{bib: Kachkachi}
\begin{equation}\label{NGaus}
\mathcal{F} = F + \int d\mathbf{r} [\gamma |\Psi|^4 + \nu|\Psi|^6]
\end{equation}
where $F$ is given by eq. (\ref{FreeEnergy}). In the absence of
the orbital magnetic field effect $\mathbf{D} = i\mathbf{\nabla}.$
Here coefficient $\gamma$ being a function of Zeeman energy and
temperature could also change sign and become negative at hight
magnetic field and low temperatures. In clean superconductors
with simple Fermi surface both coefficients
$\beta$ and $\gamma$ change sign at the same point on the
transition line - the so called tricritical point. Brazovskii \cite{bib: Brazovskii} showed that coupling of
fluctuations in inhomogeneous LOFF-like systems are important and
could lead to the first-order type transition. This is why one has to keep
term $\sim |\Psi|^6$ in GL functional.

\begin{figure}[t] \centering
\includegraphics[width=1.5cm]{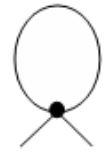}
\caption{First order fluctuation correction diagram} \label{fig:9}
\end{figure}

The first order correction in $\gamma$ is given by the bubble
containing fluctuation propagator is shown on fig. (\ref{fig:9}).
Fluctuations in LOFF-like systems are more singular near
transition. Let us consider how they depend on dimensionality of
the system. First fluctuation correction, which determines
transition temperature shift $a(T-T_{c})\rightarrow
a(T-T_{c}+\Delta T(D))$ is given by expression \cite{bib:
Patashinskii}
\begin{equation}\label{fl-cor}
a\Delta T(D) \equiv \gamma
\frac{V_D}{V}\int\frac{d^D\mathbf{p}}{(2\pi)^D}
\frac{T}{a(T-T_c)+\frac{|\beta|}{2Q^2}(p^2-Q^2)^2}
\end{equation}
Where $V/V_3=1$ for three dimensional system, $V/V_2=d$ for thin
film with thickness $d<\zeta$ and $V/V_1=S$ for thin wire with cross section area $S$ and
thickness less than $\zeta$.

Calculating (\ref{fl-cor}), we obtain at $Q\zeta >1$
\begin{equation}\label{T-shift1}
a\Delta T(D)\sim \frac{\gamma T_{c}}{|\beta|}\frac{V_D}{V}\zeta
Q^{D-1}
\end{equation}
and in the regime of small $|\beta|$ when
$Q\zeta=\sqrt{\frac{|\beta|}{2\delta}}\zeta < 1$ we estimate
\begin{equation}\label{T-shift2}
a\Delta T(D)\sim \frac{\gamma
T_{c}}{\delta}\frac{V_D}{V}\left(\frac{\zeta}{Q}\right)^{\frac{4-D}{2}}
\end{equation}

Comparing temperature shift (\ref{T-shift1}, \ref{T-shift2}) with
$T-T_{c}$, we obtain Levanuk - Ginzburg parameter $\tau_{LG}$,
which determined the width of critical fluctuations region where
gaussian approximation fails.

In the case $Q\zeta >1$ the Levanuk - Ginzburg parameter can be
estimated in different dimensions as
\begin{eqnarray}\label{crit1} \nonumber
\tau_{LG} &\sim& \left(\frac{T_c}{E_F}\right)^{4/3}G_3, D=3\\
\nonumber \tau_{LG} &\sim&
\left(\frac{1}{dp_F}\frac{T_c}{E_F}\right)^{2/3}G_2, D=2\\
\tau_{LG} &\sim& (Sp_F^2)^{-2/3}G_1, D=1
\end{eqnarray}
In estimation for quasi one dimensional wire $p_{F}$ is fermi
momentum.

In expression (\ref{crit1}) $ G_D =
\left[\frac{|\gamma|}{|\gamma_0|} \right]^{2/3}
\left[\frac{|\beta|}{|\beta_0|}\right]^{(D-2)/3}$, $\beta_0\sim
1/m$ and $\gamma_0\sim T_{c}^{2}/nE_F $ are the values of the
coefficients $\beta$ and $\gamma$ far from the LOFF - metal
transition. $m$ and $n$ are the electron mass and the density,
correspondingly. Thus one always has $|\frac{\beta_0}{\beta}|>1$
and $|\frac{\gamma_0}{\gamma}|>1$ and $G_D $ is small parameter.
In obtaining (\ref{crit1}) we use estimation for $a\sim T_c/E_F$.

In the opposite case when $Q\zeta < 1$ the Levanuk - Ginzburg
parameter can be estimated as
\begin{eqnarray}\label{crit2}\nonumber
\tau_{LG} &\sim& \left(\frac{T_c}{E_F}\right)^{4/5}\tilde{G}_3, D=3\\
\nonumber \tau_{LG} &\sim&
\left(\frac{1}{dp_F}\frac{T_c}{E_F}\right)^{2/3} \tilde{G}_2,D=2\\
\tau_{LG} &\sim& (Sp_F^2)^{-4/7} \tilde{G}_1,D=1
\end{eqnarray}
where now $\tilde{G}_D =
\left[\frac{\gamma}{\gamma_0}\right]^{4/(8-D)}$.

Indeed in three and two dimensional systems correction is much
more singular and corresponding critical region is much larger
than in case of uniform order parameter. In quasi 2D organic
superconductors and in heavy - fermion $\mathrm{CeCoIn_5}$
compound the critical fluctuations region is still very small
provided $T_c/E_F \sim 10^{-2}-10^{-3}$ \cite{bib: OS1} and
$T_c/E_F \sim 0.15$ \cite{bib: LOFF1} correspondingly.

However in one dimensional case correction coincides with that for
the case of uniform order parameter. In this sense the regimes of
quasi zero dimensional ring ($R\sim \zeta$) and quasi one
dimensional cylinder ($L>> R, \zeta$) considered here are the same
as in case of standard superconductors and do not deserve special
discussion. Parameter, which determines smallness of the Levanuk -
Ginzburg parameter in quasi one dimensional wire, is
$Sp_{F}^{2}>>1$, i.e. large number of transverse modes.

If coefficient at term $|\Psi|^4$ changes sign and becomes
negative then one should check if the first order type transition
destroys the Gaussian approximation. Let us estimate the
temperature width of first order type transition in case of
supercooling.

Consider sample with size less or order of $\zeta$. Varying
(\ref{NGaus}) with respect to the amplitude of the order parameter
written as $\Psi (\textbf{r})=\Psi \cos(\textbf{Q}\textbf{r})$ we
obtain three solutions: $\Psi_0 = 0$ and
\begin{equation}
\Psi_{\pm} =\frac{|\gamma|}{3\nu} \pm
\sqrt{\left(\frac{\gamma}{3\nu}\right)^2 -\frac{a(T-T_c)}{3\nu}}
\end{equation}

Temperature of first order transition from $\Psi_0 = 0$ state to
$\Psi_{+}$ state is determined by the condition
\begin{equation}
\mathcal{F}(\Psi_+)=\mathcal{F}(\Psi_0)=0
\end{equation}
and is larger than $T_c$.

Solution $\Psi_{-}$ corresponds to the maximum of functional
(\ref{NGaus})
\begin{equation}
\mathcal{F}(\Psi_-) \sim V \frac{(a(T-T_c))^2}{4|\gamma|}
\end{equation}

The probability of thermal activation transition of the order
parameter from the steady state $\Psi_0$ over the barrier of
height $\mathcal{F}(\Psi_-)$ is proportional to the value
\begin{equation}
\sim \exp{\left(-\frac{\mathcal{F}(\Psi_-)}{T}\right)}
\end{equation}

In case of $\mathcal{F}(\Psi_-)\gg T_c$ system will stay in
supercooled state. Corresponding temperature region might be
estimated as

\begin{equation}
(T-T_c)/T_c >
\frac{1}{aT_c}\left(\frac{T_c|\gamma|}{V}\right)^{1/2}\sim
(N_{0}VT_{c})^{-1/2}
\end{equation}

Here $N_{0}$ is electron density of states at Fermi level. In case
of not too small sample, when $N_{0}VT_{c}>>1$, supercooled
$\Psi_0 = 0$ state might be extended at $(T-T_c)/T_c < 1$.

\section{Summary}
To summarize, we showed that Aharonov - Bohm effect is very
sensitive tool for studying the intrinsic properties of
superconductors in the regime of inhomogeneous LOFF state.

Depending on the ratio of modulation period of superconducting
order parameter and the radius of the ring/cylinder transition
temperature in magnetic flux can be either increased or decreased,
or even can have double-peak structure at one flux quantum.

These effects arise due to the degeneracy of fluctuation energy
spectrum of nonhomogeneous LOFF state.

We calculated the fluctuation contribution for the persistent
current in thin superconducting ring and presented numerical
results for fluctuation specific heat and persistent current for
the cylinder. Flux dependencies of persistent current and specific
heat qualitatively correspond to that of transition temperature.

We showed that despite the enhancing of singularity due to
fluctuation's interaction in higher dimensions, in quasi one
dimensional system Livanuk-Ginzburg parameter coincides with that
for homogeneous superconductor.

Most studied reason for inhomogeneity is zeeman splitting due to
external magnetic field or exchange splitting in ferromagnetic
superconductor. In this case coefficients $\beta$, $\delta$ and
temperature $\tilde{T}_c$ itself depend on the magnetic field.
Thus AB oscillations are superimposed by monotonous dependence on
magnetic field. In ring geometry these two could be separated by
measuring of AB oscillations in tilted magnetic field.
\section{Acknowledgments}
We are grateful for the financial support of Dynasty foundation
and RFFI Grant 09-02-00571.

\end{document}